\documentclass[manuscript]{aastex} 
\usepackage{spr-astr-addons}
\usepackage{hyperref}
\usepackage{url}
\usepackage{lscape}

\shorttitle{Models of compact stars}
\shortauthors{Pandya \and Thomas}
\begin{document}
\title{Models of compact stars on paraboloidal spacetime satisfying Karmarkar condition}
\author{D. M. Pandya} 
\affil{Department of Mathematics, Pandit Deendayal Petroleum University, Raisan, Gandhinagar 382 007, India}
\email{dishantpandya777@gmail.com}
\author{V. O. Thomas}
\affil{Department of Mathematics, Faculty of Science, The Maharaja Sayajirao University of Baroda, Vadodara 390 002, India}
\email{votmsu@gmail.com}
\begin{abstract}
A new exact solution of Einstein's field equations on the background of paraboloidal spacetime using Karmarkar condition is reported. The physical acceptability conditions of the model are investigated and found that the model is compatible with a number of compact star candidates like Her X-1, LMC X-4, EXO 1785-248, PSR J1903+327, Vela X-1 and PSR J1614-2230. A noteworthy feature of the model is that it is geometrically significant and simple in form.
\end{abstract}


\section{\label{sec1}Introduction}
Ever since Schwarzschild obtained exact solution of EFEs, a wide variety of exact solutions with physical significance and devoid of any physical significance were given by a number of researchers. The analysis of solutions for physical significance revealed that out of 127 exact solutions, only 16 could withstand the elementary test for physical acceptability of the solutions (\cite{Delgaty98}). By the discovery of superdense stars like neutron stars and pulsars a new interest has emerged among researchers for developing mathematical models of such distributions. It has been suggested, theoretically, by \cite{Ruderman72} and \cite{Canuto74} that, stars, whose density in the range greater than $ 10^{15} gm/cm^3 $ may develop pressure anisotropy within it. \\
\indent \cite{Bowers74} has discussed diverse reasons for the occurrence of anisotropy inside the star. They have shown that anisotropy can affect the maximum equilibrium mass and surface redshift of the distribution. Since then, a number of anisotropic models of superdense stars have been developed and investigated (\cite{Maharaj89,Gokhroo94,Patel95,Tikekar98,Tikekar99,Tikekar05,Thomas05,Thomas07}). Impacts of anisotropy on the stability of a stellar configuration have been studied by \cite{Dev02,Dev03,Dev04}. \cite{Sharma07} and \cite{Thirukkanesh08} have obtained analytic solutions of compact anisotropic stars by assuming a linear equation of state(EOS). To solve the Einstein-Maxwell system, \cite{Komathiraj07} have used a linear equation of state. By assuming a linear EOS, \cite{Sunzu14} have reported solutions  for a charged anisotropic quark star. \cite{Feroze11} and \cite{Maharaj12} have used a quadratic-type EOS for obtaining solutions of anisotropic distributions. \cite{Varela10} have analyzed charged anisotropic configurations admitting a linear as well as non-linear equations of state. For a star composed of quark matter in the MIT bag model, \cite{Paul11} have shown how anisotropy could effect the value of the Bag constant. For a specific polytropic index, exact solutions to Einstein's field equations for an anisotropic sphere admitting a polytropic EOS have been obtained by \cite{Thirukkanesh12}. \cite{Maharaj13b} have used the same type of EOS to develop an analytical model describing a charged anisotropic sphere. \\
\indent \cite{Bhar15a,Bhar15b} and \cite{NewtonSingh15a} have shown that pressure anisotropy leads to arbitrarily large red-shifts. There has been a renewed interest among researchers to develop spacetime metrics of stellar objects of embedding class one type spacetimes (\cite{Karmarkar48}). Solutions representing superdense stars of embedding class one which are compatible with observational data of pulsars have been given by \cite{Bhar16,Singh16,Singh17,Singh17b,Singh17c}. \\
\indent In this paper we have obtained solutions of EFEs satisfying Karmarkar condition on a paraboloidal spacetime compatible with observational data of a number of compact star candidates like Her X-1, LMC X-4, EXO 1785-248, PSR J1903+327, Vela X-1 and PSR J1614-2230. The paper has been organized follows: In section \ref{sec2}, the paraboloidal spacetime metric has been discussed. Section \ref{sec3} refers to the field equations and Karmarkar condition for a static spherically symmetric anisotropic fluid sphere. In section \ref{sec4} we have solved the relevant field equations. In section \ref{sec5}, we have obtained the constants of integration $ A $ and $ B $ using the boundary conditions and displayed the explicit expressions for density $ \rho $, radial pressure $ p_r $ and anisotropy $ \Delta $. All the physical acceptability conditions have been extensively discussed in section \ref{sec6}. The expressions for the variation of physical parameters $ (\frac{d\rho}{dr}, \frac{dp_r}{dr}, \frac{dp_\perp}{dr}, \frac{d\Delta}{dr}, \frac{dp_r}{d\rho}, \frac{dp_\perp}{d\rho}, \frac{d\Delta}{d\rho}) $ is shown in section \ref{sec7}. In section \ref{sec8}, we have discussed the physical viability of the model using graphical method for the compact stars like Her X-1, LMC X-4, EXO 1785-248, PSR J1903+327, Vela X-1 and PSR J1614-2230. In section \ref{sec9}, we have concluded by pointing out the main results of our model. \\
\section{\label{sec2} The Paraboloidal Spacetime Metric}
The cartesian equation 
\begin{equation}
 x^2 + y^2 + z^2 = 2\omega R  
\end{equation}
represents a 3-paraboloid immersed in a 4-dimensional Euclidean Space.
Here $ x = const,~y=const $ and $ z = const $ represent three-paraboloids while $ \omega = const $ are spheres.
Under the parametrization
\begin{eqnarray}
 \nonumber x = r~sin\theta~cos\phi, \\
 \nonumber y = r~sin\theta~cos\phi, \\
 \nonumber z = r~cos\theta, \\
 \omega = \frac{r^2}{2R},  
\end{eqnarray}
the Euclidean metric 
\begin{equation}
 d\sigma^2 = dx^2 + dy^2 + dz^2 + d\omega^2
\end{equation}
becomes
\begin{equation}
 d\sigma^2 = \left(1+\frac{r^2}{R^2}\right)dr^2 + r^2 d\theta^2 + r^2 sin^2\theta~d\phi^2,
\end{equation}
where $ R $ is a geometric parameter. \\
We shall consider the spacetime metric 
\begin{mathletters}
\begin{eqnarray}
 \label{1} ds^2 = e^{\nu(r)}dt^2 - e^{\lambda}dr^2 - r^2 d\theta^2 - r^2 sin^2\theta~d\phi^2 \\
 \label{2} \mbox{with}~~~e^{\lambda} = \left(1 + \frac{r^2}{R^2} \right)
\end{eqnarray}
\end{mathletters}
for describing the interior of anisotropic fluid distribution. A detailed study of the metric (5) has been done by \cite{Jotania06} \\
\section{\label{sec3} The Field Equations and Karmarkar Condition}
The energy-momentum tensor for anisotropic matter distribution is taken as 
\begin{equation}
 T_i^j = (\rho + p_r)u^ju_i - p_\perp \delta_i^j + (p_r - p_\perp)\eta^j\eta_i
 \label{E_M_Tensor}
\end{equation}
where $ u_i $ denotes the four-velocity and $ \eta_i $ is a space like vector orthogonal to $ u^i $ satisfying the conditions 
\begin{equation}
  u^ju_i = 1,~\eta^j\eta_i = -1~\mbox{and}~u^j\eta_i = 0.
\end{equation}
$ \rho,p_r,p_\perp $ denotes the proper density, the radial pressure and the transverse pressure, respectively.\\
The Einstein's field equations for the metric (\ref{1}) with energy-momentum tensor (\ref{E_M_Tensor}) are equivalent to the following set of three equations
\begin{eqnarray}
 \label{rhoefe} 8\pi\rho = \frac{e^{-\lambda} \lambda'}{r} + \frac{1 - e^{-\lambda}}{r^2}, \\
 \label{radialefe} 8\pi p_r = \frac{e^{-\lambda} \nu'}{r} + \frac{e^{-\lambda}-1}{r^2}, \\
 \label{pperpefe} 8\pi p_\perp = e^{-\lambda} \left(\frac{\nu''}{2} + \frac{\nu'^2}{4} - \frac{\nu'\lambda'}{4} + \frac{\nu'-\lambda'}{2r}\right).  	
\end{eqnarray}
Equations (\ref{rhoefe}) -- (\ref{pperpefe}) consist of a system of three equations in five unknowns $ (\lambda,\nu,\rho,p_r, p_\perp) $. One of the variables $ \lambda $ is known from (\ref{2}). Once we know the value of $ \nu $, the values of $ \rho,p_r,p_\perp $ can be obtained from equations (\ref{rhoefe}), (\ref{radialefe}) and (\ref{pperpefe}). \\
The spacetime metric (\ref{1}) is of class one type if it satisfies the Karmarkar condition \citep{Karmarkar48} given by
\begin{equation}
 R_{1414}R_{2323} = R_{1212}R_{3434} + R_{1224}R_{1334}
 \label{Karmarkar_Condition}
\end{equation}
with $ R_{2323} \neq 0 $ where components of Riemann curvature tensor are given by 
\begin{eqnarray*}
 R_{2323}=r^2 sin^2\theta \left[1 - e^{-\lambda}\right], \\
 R_{1212} = \frac{1}{2} \lambda' r, \\
 R_{1334} = R_{1224}~sin^2 \theta = 0, \\
 R_{1414} = -e^{\nu} \left[ \frac{\nu''}{2} + \frac{\nu'^2}{4} - \frac{1}{4} \lambda'\nu'\right], \\
 R_{2424} = -\frac{1}{4} \nu'r e^{\nu - \lambda}, \\
 R_{3434} = sin^2\theta R_{2424}.
\end{eqnarray*}
The Karmarkar condition (\ref{Karmarkar_Condition}) leads to the differential equation 
\begin{equation}
 \frac{2\nu''}{\nu'} + \nu' = \frac{\lambda' e^\lambda}{e^{\lambda}-1}
 \label{Kar_Diff_Eqn}
\end{equation}
Using the expression of $ e^{\lambda} $ given in (\ref{2}), equation (\ref{Kar_Diff_Eqn}) becomes 
\begin{equation}
 \frac{2\nu''}{\nu'} + \nu'= \frac{2}{r}
\end{equation}
which gives a closed form solution
\begin{equation}
 e^{\nu} = \left(A + B \frac{r^2}{R^2}\right)^2.
 \label{eraisedtonu}
\end{equation}
The explicit form of the spacetime metric is 
\begin{eqnarray}
 \nonumber ds^2 = \left(A + B \frac{r^2}{R^2}\right)^2 dt^2 - \left(1 + \frac{r^2}{R^2}\right) dr^2 - r^2 d\theta^2 \\
 - r^2 sin^2\theta~d\phi^2,
\end{eqnarray}
where $ A $ and $ B $ are constants of integration which are to be determined using appropriate boundary conditions.\\
\section{\label{sec4}Solution of Field Equations}
The field equations (\ref{rhoefe}) -- (\ref{pperpefe}) can now be solved by using the values of $ \lambda $ and $ \nu $ given by equations (\ref{2}) and (\ref{eraisedtonu}). The expressions for $ \rho, p_r $ and $ p_\perp $ and the anisotropy factor $ \Delta (= p_r - p_\perp) $ are given, respectively, by 
\begin{eqnarray}
 8\pi\rho = \frac{3 + \frac{r^2}{R^2}}{R^2 \left(1 + \frac{r^2}{R^2}\right)^2}, \label{rhonew}\\
 8\pi p_r = \frac{B \left(4 - \frac{r^2}{R^2}\right) - A}{R^2 \left(A + B \frac{r^2}{R^2}\right)\left(1 + \frac{r^2}{R^2}\right)}, \label{prnew} \\
 8\pi p_\perp = \frac{B \left(4 + \frac{r^2}{R^2}\right)-A}{R^2 \left(A + B \frac{r^2}{R^2}\right)\left(1 + \frac{r^2}{R^2}\right)^2},
 \label{pperpnew}
\end{eqnarray}
and
\begin{equation}
 8\pi\Delta = \frac{\frac{r^2}{R^2}\left[B\left(2-\frac{r^2}{R^2}\right)-A\right]}{R^2 \left(A + B \frac{r^2}{R^2}\right) \left(1+\frac{r^2}{R^2}\right)^2}
 \label{anisotropynew}
\end{equation}
The anisotropy $ \Delta $ vanishes at $ r = 0, $ which is a required condition. \\
\section{\label{sec5}Boundary Conditions}
The interior spacetime metric (5) should match continuously with the Schwarzschild exterior metric
\begin{eqnarray}
 \nonumber ds^2 = \left(1 - \frac{2M}{r}\right) dt^2 - \frac{1}{1-\frac{2M}{r}}dr^2 - r^2 d\theta^2 \\
 - r^2 sin^2\theta~d\phi^2,
 \label{Sch_Ext_Metric}
\end{eqnarray}
across the boundary $ r = a $. This gives 
\begin{equation}
 1 - \frac{2M}{a} = \frac{1}{1 + \frac{a^2}{R^2}}
\end{equation}
determining the values of the geometric parameter $ R $ in terms of $ a $ and $ M $ by the relation 
\begin{equation}
 R = a \sqrt{\frac{a}{2M}-1} \label{radius}.
\end{equation}
The total mass enclosed within the radius $ a $ is given by 
\begin{equation}
 M = \frac{\frac{a^3}{R^2}}{2\left(1 + \frac{a^2}{R^2}\right)}
\end{equation}
Equating the coefficients of $ dt^2, $ we get 
\begin{equation}
 1 - \frac{2M}{a} = \left(A + B \frac{r^2}{R^2}\right)^2 = \frac{1}{1 + \frac{a^2}{R^2}}
\end{equation}
which gives
\begin{equation}
 A + B \frac{a^2}{R^2}  = \frac{1}{\sqrt{1 + \frac{a^2}{R^2}}}
\label{bdrycondn1}
\end{equation}
The second boundary condition is given by $ p_r (r = a) = 0. $ This leads to
\begin{equation}
 - A + B \left(4 - \frac{a^2}{R^2}\right) = 0.
 \label{bdrycondn2}
\end{equation}
Equations (\ref{bdrycondn1}) and (\ref{bdrycondn2}) determine the values of $ A $ and $ B $ in the form 
\begin{eqnarray}
 A = \frac{4-\frac{a^2}{R^2}}{4\sqrt{1+\frac{a^2}{R^2}}},
 \label{A} \\
 B = \frac{1}{4\sqrt{1+\frac{a^2}{R^2}}}.
 \label{B}
\end{eqnarray}
Using (\ref{A}) and (\ref{B}) we rewrite equations (\ref{prnew}) -- (\ref{anisotropynew}) as 
\begin{eqnarray}
 8\pi p_r = \frac{\frac{a^2}{R^2} - \frac{r^2}{R^2}}{R^2 \left(4+\frac{r^2}{R^2}-\frac{a^2}{R^2}\right)\left(1+\frac{r^2}{R^2}\right)},
 \label{prfinal} \\
 8\pi p_\perp = \frac{\frac{a^2}{R^2} + \frac{r^2}{R^2}}{R^2 \left(4+\frac{r^2}{R^2}-\frac{a^2}{R^2}\right)\left(1+\frac{r^2}{R^2}\right)^2},
 \label{pperpfinal} \\
 8 \pi \Delta = \frac{\frac{r^2}{R^2}\left(\frac{a^2}{R^2} - \frac{r^2}{R^2} - 2\right)}{R^2 \left(4+\frac{r^2}{R^2}-\frac{a^2}{R^2}\right)\left(1+\frac{r^2}{R^2}\right)^2}.
 \label{anisotropyfinal}
\end{eqnarray}
\section{\label{sec6}Physical Acceptability Conditions}
A physically acceptable anisotropic stellar model must satisfy the following conditions (\cite{Kuchowicz72,Buchdahl79,Murad15,Knutsen87}): \\
\renewcommand\theenumi{\textbf{(\alph{enumi})}}
\renewcommand\theenumii{\roman{enumii}}
\begin{enumerate}
 \item Regularity conditions
\begin{enumerate}
 \item The metric potentials $ e^{\lambda} > 0,~e^{\nu} > 0, $ for $ 0 \leq r \leq a $. \\ 
 \item $ \rho(r) \geq 0,~p_r(r) \geq 0, p_\perp(r) \geq 0 $ for $ 0 \leq r \leq a. $ \\
 \item $ p_r (r = a) = 0 $.\\
\end{enumerate}
\item Causality conditions
\begin{enumerate}
 \item $ 0 \leq \frac{dp_r}{d\rho} \leq 1 $, for $ 0 \leq r \leq a $. \\
 \item $ 0 \leq \frac{dp_\perp}{d\rho} \leq 1 $, for $ 0 \leq r \leq a $. \\ 
\end{enumerate}
\item Energy conditions
\begin{enumerate}
 \item $ \rho - p_r  - 2 p_\perp \geq 0 $ (strong energy condition),
 \item $ \rho \geq p_r,~\rho \geq p_\perp $ (weak energy conditions) \\
\end{enumerate}
\item Monotone decrease of physical parameters
\begin{enumerate}
 \item $ \frac{d\rho}{dr} \leq 0,~\frac{dp_r}{d\rho} \leq 0 $ for $ o \leq r \leq a, $ \\
 \item $ \frac{d}{dr} \left(\frac{dp_r}{d\rho}\right) \leq 0, $ for $ 0 \leq r \leq a, $ \\
 \item $ \frac{d}{dr} \left(\frac{p_r}{\rho}\right) \leq 0 $ for $ 0 \leq r \leq a. $ \\
\end{enumerate}
\item Pressure anisotropy \\
$ \Delta (r = 0) = 0. $ \\
\item Mass-radius relation \\
According to \cite{Buchdahl79}, the allowable mass radius relation must satisfy the inequality $ \frac{M}{R} \leq \frac{4}{9}. $\\
\item Redshift \\
The redshift $ z = e^{-\frac{\nu}{2}}-1 $ must be a decreasing function of $ r $ and finite for $ 0 \leq r \leq a. $\\
\item Stability condition \\
 The relativistic adiabatic index $ \Gamma = \frac{\rho + p_r}{p_r} \frac{dp_r}{d\rho} \geq \frac{4}{3} $ for $ 0 \leq r \leq R. $
\end{enumerate}
\section{\label{sec7}Variation of Physical Parameters}
The variation of density $ \rho $ with respect to the radial variable $ r $ is given by 
\begin{equation}
 8\pi \frac{d\rho}{dr} = - \frac{2}{R^2}\cdot\frac{r}{R^2} \frac{\left(5 + \frac{r^2}{R^2}\right)}{\left(1 + \frac{r^2}{R^2}\right)^3} 
\end{equation}
Since $ \frac{d\rho}{dr} < 0, $ for $ 0 < r \leq a, $ the density distribution decreases radially outward.\\
The gradient of radial pressure, transverse pressure and the anisotropy variable have the following expressions
\begin{eqnarray}
8\pi \frac{dp_r}{dr} = -\frac{2 r}{R^4} \frac{4 \left(1 + \frac{r^2}{R^2}\right) + \left(\frac{a^2}{R^2} - \frac{r^2}{R^2}\right) \left(4 + \frac{r^2}{R^2} - \frac{a^2}{R^2} \right)}{\left(4 + \frac{r^2}{R^2} - \frac{a^2}{R^2}\right)^2\left(1 + \frac{r^2}{R^2}\right)^2}
 \label{dprbydr}
\end{eqnarray}
\begin{eqnarray}
 8 \pi \frac{dp_\perp }{dr} = \frac{2r}{R^4}\frac{\left[4\left(1 - \frac{r^2}{R^2}\right) - 2 \frac{a^2}{R^2} \left(5 + \frac{r^2}{R^2}\right) + 2 \left(\frac{a^4}{R^4} - \frac{r^4}{R^4}\right)\right]}{\left(4 + \frac{r^2}{R^2} - \frac{a^2}{R^2}\right)^2\left(1 + \frac{r^2}{R^2}\right)^3} 
 \label {dpperpbydr}
 \end{eqnarray}
\begin{eqnarray}
\nonumber 8 \pi \frac{d\Delta}{dr} = \frac{2r}{R^4} \times \frac{1}{\left(4 + \frac{r^2}{R^2} - \frac{a^2}{R^2}\right)^2\left(1 + \frac{r^2}{R^2}\right)^3} \times  \\ 
\nonumber \left\lbrace \left(4 + \frac{r^2}{R^2} - \frac{a^2}{R^2}\right) \left(\frac{a^2}{R^2} -\frac{2r^2}{R^2} - \frac{r^4}{R^4} - 2 \right) \right.\\
\left. - \frac{r^2}{R^2}\left(1 + \frac{r^2}{R^2}\right)\left(\frac{a^2}{R^2} - \frac{r^2}{R^2}-2\right) \right\rbrace
 \label{dDeltabydr}
\end{eqnarray}
It can be seen from equation (\ref{dprbydr}) that $ 8 \pi \frac{dp_r}{dr} < 0 $ for $ \frac{a^2}{R^2} < 4 $. This indicates that radial pressure is a decreasing function of $ r $. However due to the complexity of expressions in the right hand side of equations (\ref{dpperpbydr}) and (\ref{dDeltabydr}), it is difficult to obtain the sign of the terms in their right hand side. However it can be seen from equations (\ref{pperpfinal}) and (\ref{anisotropyfinal}) that $ p_\perp $ and $ \Delta $ are also decreasing functions of $ r $.\\
 The sequence of the radial and transverse speed of sound, $ \upsilon_r^2 $ and $ \upsilon_\perp^2 $, are given by
\begin{eqnarray}
\nonumber \upsilon_r^2 = \frac{dp_r}{d\rho} = \frac{\left(1 + \frac{r^2}{R^2}\right)}{\left(4 + \frac{r^2}{R^2} - \frac{a^2}{R^2}\right)^2\left(5 + \frac{r^2}{R^2}\right)} \times \\
\left[4\left(1 + \frac{r^2}{R^2}\right) + \left(\frac{a^2}{R^2} - \frac{r^2}{R^2}\right)\left(4 + \frac{r^2}{R^2} - \frac{a^2}{R^2}\right)\right]\label{dprbydrho}
\end{eqnarray}
\begin{eqnarray}
 \upsilon_\perp^2 = -\frac{4 \left(1 - \frac{r^2}{R^2}\right) - 2\frac{a^2}{R^2}\left(5 + \frac{r^2}{R^2}\right) + 2 \left(\frac{a^4}{R^4} - \frac{r^4}{R^4}\right)}{\left(4 + \frac{r^2}{R^2} - \frac{a^2}{R^2}\right)^2\left(5 + \frac{r^2}{R^2}\right)} \label{dpperpbydrho}
\end{eqnarray}
\begin{eqnarray}
\nonumber \frac{d\Delta}{d\rho} = - \frac{1}{\left(4 + \frac{r^2}{R^2} - \frac{a^2}{R^2}\right)^2\left(5 + \frac{r^2}{R^2}\right)} \times \\
\nonumber \left[\left(4 + \frac{r^2}{R^2} - \frac{a^2}{R^2}\right) \left(\frac{a^2}{R^2} - 2 \frac{r^2}{R^2} - \frac{r^4}{R^4} - 2 \right) \right. \\
\left. - \frac{r^2}{R^2}\left(1 + \frac{r^2}{R^2}\right)\left(\frac{a^2}{R^2} - \frac{r^2}{R^2} - 2 \right)\right]
\end{eqnarray}
\section{\label{sec8}Physical analysis}
\begin{figure}
\plotone{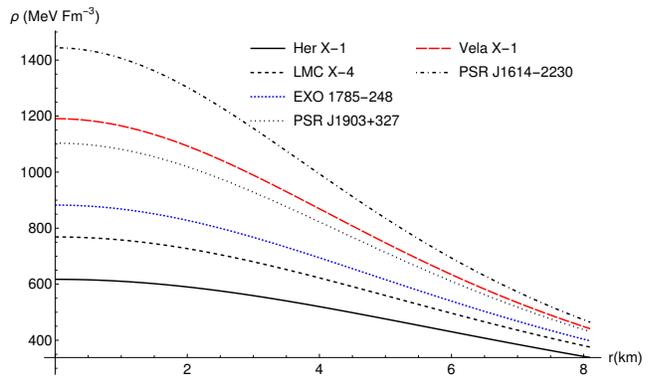}
\caption{Density profile \label{fig1}}
\end{figure}

\begin{figure}
\plotone{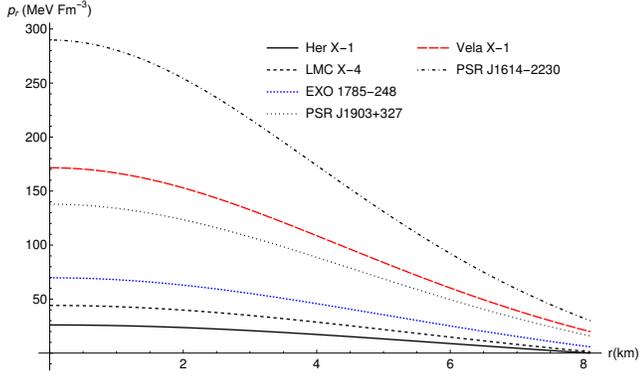}
\caption{Radial pressure profile \label{fig2}}
\end{figure}

\begin{figure}
\plotone{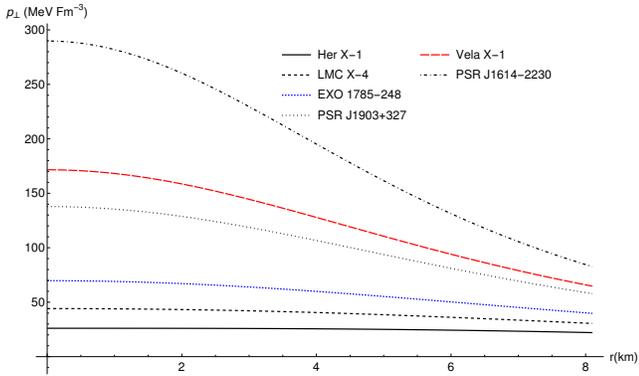}
\caption{Transverse pressure profile \label{fig3}}
\end{figure}

\begin{figure}
\plotone{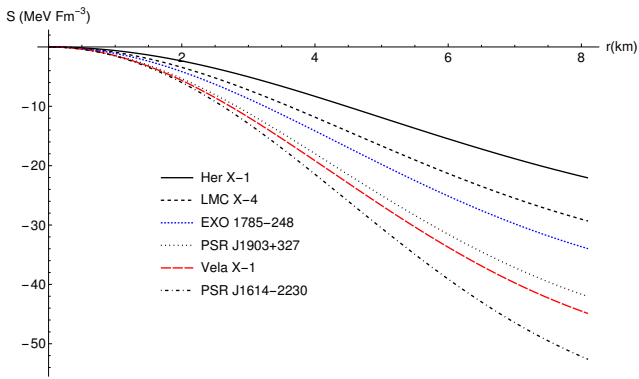}
\caption{Anisotropy profile \label{fig4}}
\end{figure}

\begin{figure}
\plotone{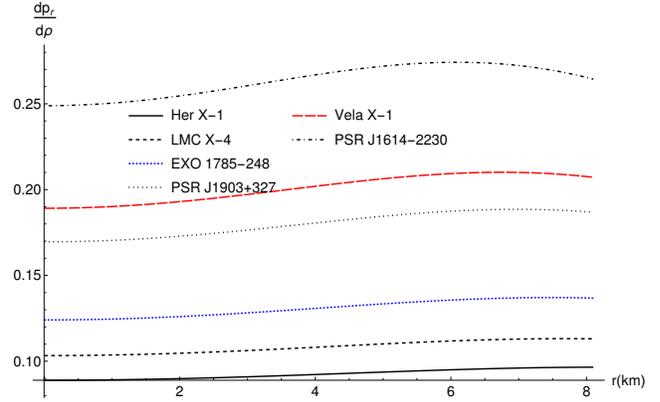}
\caption{Radial sound speed profile \label{fig5}}
\end{figure}

\begin{figure}
\plotone{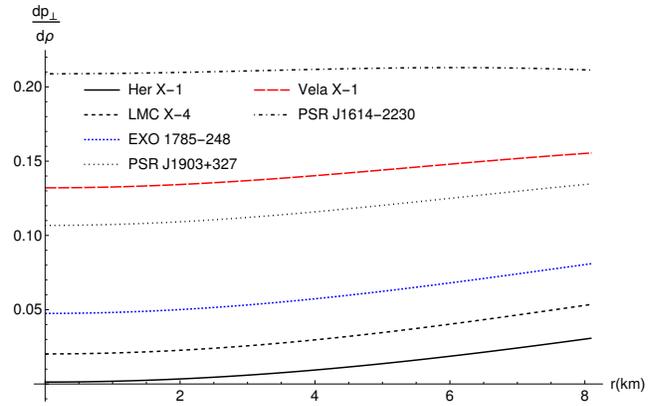}
\caption{Transverse sound speed profile \label{fig6}}
\end{figure}

\begin{figure}
\plotone{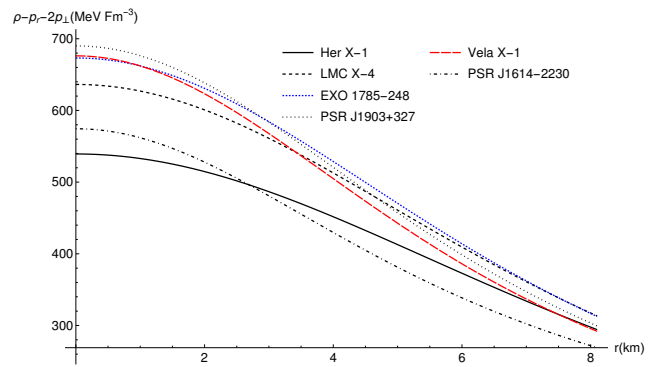}
\caption{Strong energy condition profile \label{fig8}}
\end{figure}

\begin{figure}
\plotone{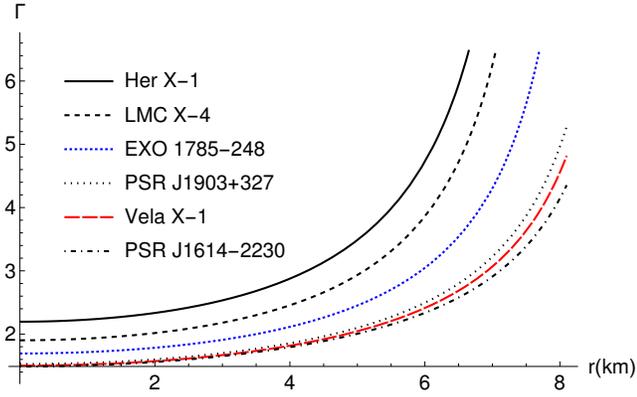}
\caption{Adiabatic index profile. \label{fig9}}
\end{figure}

\begin{figure}
\plotone{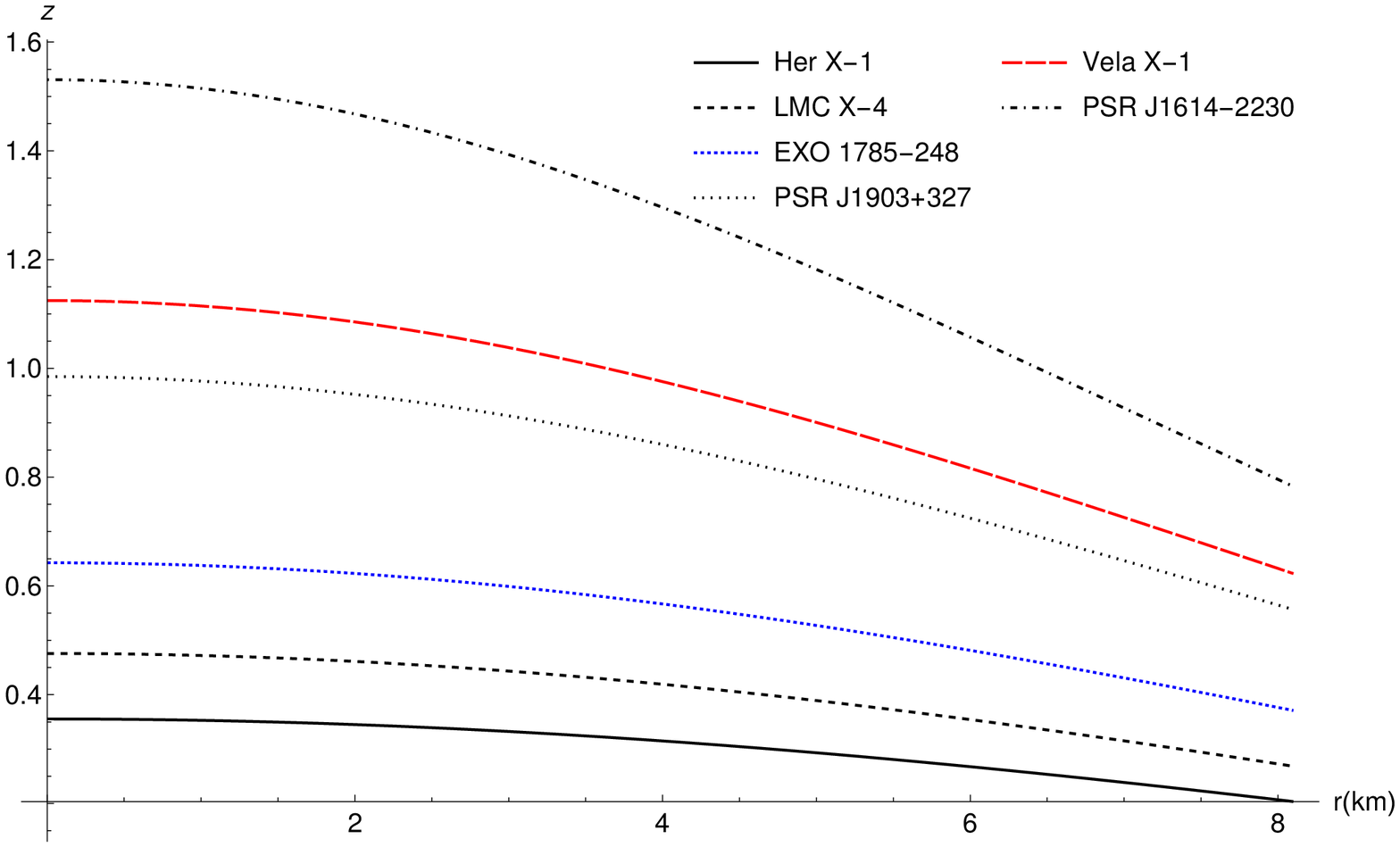}
\caption{Redshift profile. \label{fig10}}
\end{figure}

In order to examine the compatibility of the model with observational data, we have considered compact stars Her X-1, LMC X-4, EXO 1785-248, PSR J1903+327, Vela X-1 and PSR J1614-2230 whose mass and size are known \citep{Gangopadhyay13}. By taking the mass $ M $ and radius $ a $, the value of the geometric parameter $ R $ is found from equations (\ref{radius}).
\begin{table}[ht]
\centering
\scriptsize
\caption{The compactness $ u = \frac{M}{a} $ of different stars are shown in the following table.}
\label{tab:1}
\vspace{0.3cm}
 \begin{tabular}{ccccc} \hline \\
 Star & M & $ a $ & $ R $ & compactness \\
 & ($ M_{\odot} $) & (km) & (km) & \\ \hline \\
 Her X-1 & 0.85 & 8.1 & 12.096 & 0.1548 \\
 LMC X-4 & 1.04 & 8.301 & 10.8412 & 0.1848  \\
 EXO 1785-248 & 1.3 & 10 & 10.1182 & 0.1917 \\
 PSR J1903+327 & 1.667 & 9.438 & 9.048 & 0.2605 \\
 Vela X-1 & 1.77 & 9.56 & 8.71 & 0.2731 \\	
 PSR J1614-2230 & 1.97 & 9.69 & 7.91 & 0.2999 \\ \hline
 \end{tabular} 
\end{table}
In Fig.\ref{fig1} we have shown the variations of density against the radius. It can be seen that the density accommodated by a star increases as the compactness increases. Her X-1 accommodates minimum density whose compactness is minimum while PSR J1614-2230 has maximum density for which compactness is maximum among all compact star candidates studied. \\
\indent The variation of radial pressure $ p_r $ and transverse pressure $ p_\perp $ are shown in Figs.\ref{fig2} and \ref{fig3} respectively. It can be seen that both $ p_r $ and $ p_\perp $ decreases radially outward and they increase with compactness.\\
\indent In Fig.\ref{fig4}, we have shown the variation of anisotropy against radial directions. Its value is zero at the centre and $ |\Delta| $ increases radially outward. The anisotropy is negative throughout the distribution. Further, the numerical value of anisotropy is more for stars with more compactness. \\
\indent The velocity of sound in the radial direction $ \upsilon_r^2 $ and transverse direction $ \upsilon_\perp^2 $ are shown in Figs.\ref{fig5} and \ref{fig6} respectively. From both figures it can be noticed that $ 0 < \frac{dp_r}{d\rho} < 1, $ and $ 0 < \frac{dp_\perp}{d\rho} < 1. $ Further these velocities are more in magnitude in more compact stars. \\
\indent Fig.\ref{fig8} shows that the strong energy condition satisfies for all stars throughout the distribution. \\
\indent The variation of adiabatic index is displayed in Fig.\ref{fig9}. Its value is greater than $ \frac{4}{3} $ throughout and the star with less compactness accommodate more $ \Gamma $ value. It can be noticed that as compactness increases, the value of $ \Gamma $ decreases. This indicates that the star become less stable due to the increase in compactness. \\
\indent In Fig.\ref{fig10} we have shown the variation of redshift $ z $ in the radial direction. Stars with more compactness shows more redshift. That is while Her X-1 shows minimum redshift while PSR J1614-2230 shows max redshift among all the compact star candidates. \\
\section{\label{sec9}Discussion}
We have studied the compatibility of the model developed using Karmakar condition in the background of paraboloidal spacetime for compact stars like Her X-1, LMC X-4, EXO 1785-248, PSR J1903+327, Vela X-1 and PSR J1614-2230. It is found that our model satisfy the elementary physical requirements for representing a superdense compact star through graphical method. It is found that the model developed can accommodate the mass and radius of many of the compact star candidates given by \cite{Gangopadhyay13}. \\
\indent It is found that stars whose compactness is more accommodate more density, pressure and $ \Delta. $ The redshift increase with compactness while the value of $ \Gamma $ decreases with compactness showing that the stability decreases with increase in compactness. \\
\indent A pertinent feature of the model is that the exact solution obtained is simple in nature which is seldom found in many solutions. Though we have displayed here the physical analysis, only for few compact star models, it can be applied to a larger class of known pulsars. The model possesses a definite background spacetime geometry, namely paraboloidal geometry, and the expression involved in the solution are simple in nature.\\

\end{document}